\documentclass[%
 reprint,
superscriptaddress,
 amsmath,amssymb,
 aps,
prl,
floatfix
]{revtex4-2}

\pdfoutput=1
\usepackage{comment}
\usepackage{graphicx}
\usepackage{dcolumn}
\usepackage{bm}
\usepackage[version=4]{mhchem}
\usepackage{xcolor}
\usepackage{todonotes}
\usepackage{hyperref}
\usepackage{color,soul}
\hypersetup{hidelinks,backref=true,pagebackref=true,hyperindex=true,colorlinks=true,breaklinks=true,urlcolor= blue,bookmarks=true,bookmarksopen=false,pdftitle={`Title},pdfauthor={Author}} 

\begin{document}

\preprint{Spin excitations of Infinite-Layer nickelates}

\title{Spin excitations in Nd$_{1-x}$Sr$_{x}$NiO$_2$ and YBa$_2$Cu$_3$O$_{7-\delta}$:\\ the influence of Hubbard $U$}

\author{F.\,Rosa}
 \email[]{francesco1.rosa@polimi.it}
 \affiliation{Dipartimento di Fisica, Politecnico di Milano, piazza Leonardo da Vinci 32, I-20133 Milano, Italy}

\author{L.\,Martinelli}
\altaffiliation[Present address: ]{Physik-Institut, Universit\"{a}t Z\"{u}rich, Winterthurerstrasse 
190, CH-8057 Z\"{u}rich, Switzerland}
\affiliation{Dipartimento di Fisica, Politecnico di Milano, piazza Leonardo da Vinci 32, I-20133 Milano, Italy}

\author{G.\,Krieger}
\altaffiliation[Present address: ]{Eindhoven University of Technology, P.O. Box 513, 5600 MB Eindhoven, The Netherlands}
\affiliation{Université de Strasbourg, CNRS, IPCMS UMR 7504, F-67034 Strasbourg, France
}
 
\author{L.\,Braicovich}
\affiliation{
 ESRF, The European Synchrotron, 71 Avenue des Martyrs, CS 40220, F-38043 Grenoble, France
}

\author{N.B.\,Brookes}
\affiliation{
 ESRF, The European Synchrotron, 71 Avenue des Martyrs, CS 40220, F-38043 Grenoble, France
}

\author{G.\,Merzoni}
 \affiliation{Dipartimento di Fisica, Politecnico di Milano, piazza Leonardo da Vinci 32, I-20133 Milano, Italy}
 \affiliation{European XFEL, Holzkoppel 4, Schenefeld, D-22869, Germany}

\author{M.\,Moretti Sala}
\affiliation{Dipartimento di Fisica, Politecnico di Milano, piazza Leonardo da Vinci 32, I-20133 Milano, Italy}

\author{F.\,Yakhou-Harris}
\affiliation{
 ESRF, The European Synchrotron, 71 Avenue des Martyrs, CS 40220, F-38043 Grenoble, France
}

\author{R.\,Arpaia}
\affiliation{
Quantum Device Physics Laboratory, Department of Microtechnology and Nanoscience, Chalmers University of Technology, SE-41296 Göteborg, Sweden
}
\affiliation{Department of Molecular Sciences and Nanosystems, Ca’ Foscari University of Venice, I-30172 Venice, Italy}

\author{D.\,Preziosi}
\affiliation{Université de Strasbourg, CNRS, IPCMS UMR 7504, F-67034 Strasbourg, France
}

\author{M.\,Salluzzo}
\affiliation{CNR-SPIN, Complesso Monte Sant’Angelo-Via Cinthia, I-80126 Napoli, Italy
}

\author{M.\,Fidrysiak}
\affiliation{Institute of Theoretical Physics, Jagiellonian University, ul. Łojasiewicza 11, PL-30348 Kraków, Poland
}

\author{G.\,Ghiringhelli}
\email[]{giacomo.ghiringhelli@polimi.it}
\affiliation{Dipartimento di Fisica, Politecnico di Milano, piazza Leonardo da Vinci 32, I-20133 Milano, Italy}
\affiliation{CNR-SPIN, Dipartimento di Fisica, Politecnico di Milano, I-20133 Milano, Italy}

\date{\today}

\begin{abstract} 
We use Resonant Inelastic X-ray Scattering (RIXS) to compare the doping dependence of magnetic excitations of an Infinite-Layer nickelate to those of a prototypical superconducting cuprate. 
The polarization analysis of RIXS spectra establishes the dominant spin-flip nature of the mid-infrared peak in both cases. Hole doping leads to opposite behavior of the magnetic energy in the two materials. By fitting the data with an original Hubbard-based model for dynamic susceptibility, we find that $t$ is comparable in the two materials while $U$ is about twice larger in the nickelate. This finding accounts for the smaller magnetic bandwidth of nickelates and for its decrease upon doping.

\begin{description}
\item[DOI]

\end{description}
\end{abstract}

\maketitle





\emph{Introduction} -- In high-$T_\textbf{c}$ superconducting cuprates the ground state and the transport properties are determined by an intricate 
tangle of orbital, charge, spin and lattice degrees of freedom \cite{keimer2015quantum,Scalapino,Fradkin,proust2019remarkable,weber2012scaling, foley2019coexistence,zaanen1985band,zaanen1990systematics, wahlberg2021restored}. Consequently, a consensus on how to properly describe their superconducting (SC) and normal state properties is still missing despite three decades of experimental and theoretical efforts. The problem can be attacked indirectly by replicating the physics of cuprates in other materials. Infinite-Layer (IL) nickelates are particularly well suited to this task \cite{nomura2022superconductivity,been2021electronic,botana2017electron,anisimov1999electronic}. Indeed, the $3d^9$ spin-$\frac{1}{2}$ square-lattice of Cu$^{2+}$ ions is mimicked by  Ni$^{1+}$ ions in R$_{(1-x)}$Sr$_{x}$NiO$_2$ (R stands of La, Nd or Pr), leading to superconductivity \cite{li2019superconductivity} and to electronic properties that can be conveniently compared to cuprates.
\\The planar square coordination is required for the chemical stabilization of the monovalent Ni$^{1+}$ ion \cite{anisimov1999electronic,botana2020similarities, hepting2020electronic,rossi2021orbital} and ensures that, by removing the degeneracy between the two $e_g$ states, the Ni$^{2+}$ hole-doped sites are in a low-spin configuration compatible with superconductivity \cite{jiang2020critical}. This quasi-2D structure can only be achieved in infinite-layer ultra-thin films, obtained by de-intercalating apical oxygen atoms from pristine perovskite structure \cite{preziosi2017reproducibility,lee2020aspects,ding2023critical,puphal2023synthesis,onozuka2016formation,crespin1983reduced,hayward1999sodium,hayward2003synthesis}. Notwithstanding several affinities in their temperature-doping ($T/p$) phase diagrams \cite{zeng2020phase,li2020superconducting}, IL nickelates show some remarkable differences \cite{lee2004infinite} with respect to cuprates. Firstly, the charge-transfer energy $\Delta$ is larger, therefore confining doping holes on Ni $3d$ orbitals \cite{goodge2021doping,rossi2021orbital,jiang2020critical}. 
Secondly, the rare-earth atom plays a more significant role than in cuprates \cite{jiang2019electronic,kapeghian2020electronic,hepting2020electronic}, providing pockets of $5d$ states close to the Fermi level. While these $5d$ electron pockets are little hybridized with Ni $3d_{x^2-y^2}$ states \cite{osada2021nickelate}, thus minimally affecting antiferromagnetic correlations in the NiO$_2$ planes, they can provide self-doped holes even in the undoped RNiO$_2$ \cite{zhang2020self}. This self-doping might explain the report of  superconductivity in nominally undoped LaNiO$_2$ \cite{osada2021nickelate}.
\\Muon-spin rotation measurements provided evidence of coexisting magnetism and SC \cite{fowlie2022intrinsic}. Moreover, the dispersion of spin excitations measured by Resonant Inelastic X-ray Scattering (RIXS) indicated the robust settling of antiferromagnetic correlations in the NiO$_2$ planes, although the characteristic energy of damped magnons (paramagnons) softens as a function of doping \cite{lu2021magnetic, krieger2022charge}, contrary to cuprates \cite{jia2014Persistenta}. This softening was attributed to the spin dilution effect, i.e., a decrease in the average spin moment due to the introduction of doped holes on Ni sites. However, a clear description of the phenomenon was hindered by the difficulty in disentangling the coherent spin response from electron-hole excitations. 
\\In this letter, 
by using polarization-resolved RIXS we are able to single-out magnetic and charge excitations in the low energy region of undoped and Sr-doped NdNiO$_2$ and of antiferromagnetic and Ca-doped YBa$_2$Cu$_3$O$_{7-\delta}$ (YBCO). We confirm the magnetic nature of the mid-infrared excitation in IL-nickelates and the softening of paramagnons in doped samples. Besides the usual damped harmonic oscillator fit, we analyze the spectral shape with a Hubbard-type theoretical model allowing the determination of both the hopping integral $t$ and the Coulomb repulsion $U$ \cite{kitatani2020nickelate,been2021electronic,worm2022correlations}. We demonstrate that a smaller $t/U$ ratio can explain the narrower magnetic bandwidth of nickelates with respect to cuprates even without invoking spin dilution.  

\begin{figure}[tb]
\centering
\includegraphics[width=1\columnwidth]{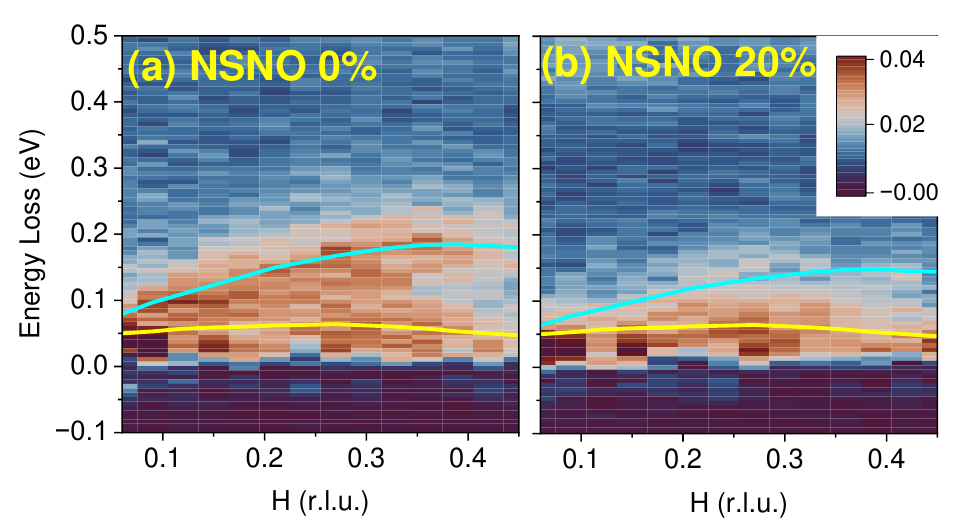}
\caption{\label{NSNO_maps} 
Energy loss/momentum maps of Nd$_{1-x}$Sr$_{x}$NiO$_2$, for the two doping levels $x=0$ (a) and $x=0.2$  (b). The intensity scale is the same, shown in the (b) inset. The elastic peak has been subtracted for convenience. Yellow and cyan lines are eye guides for the phonon and magnon peak respectively. 
} 
\end{figure}

\begin{figure*}[tb]
\centering
\includegraphics[width=\textwidth]{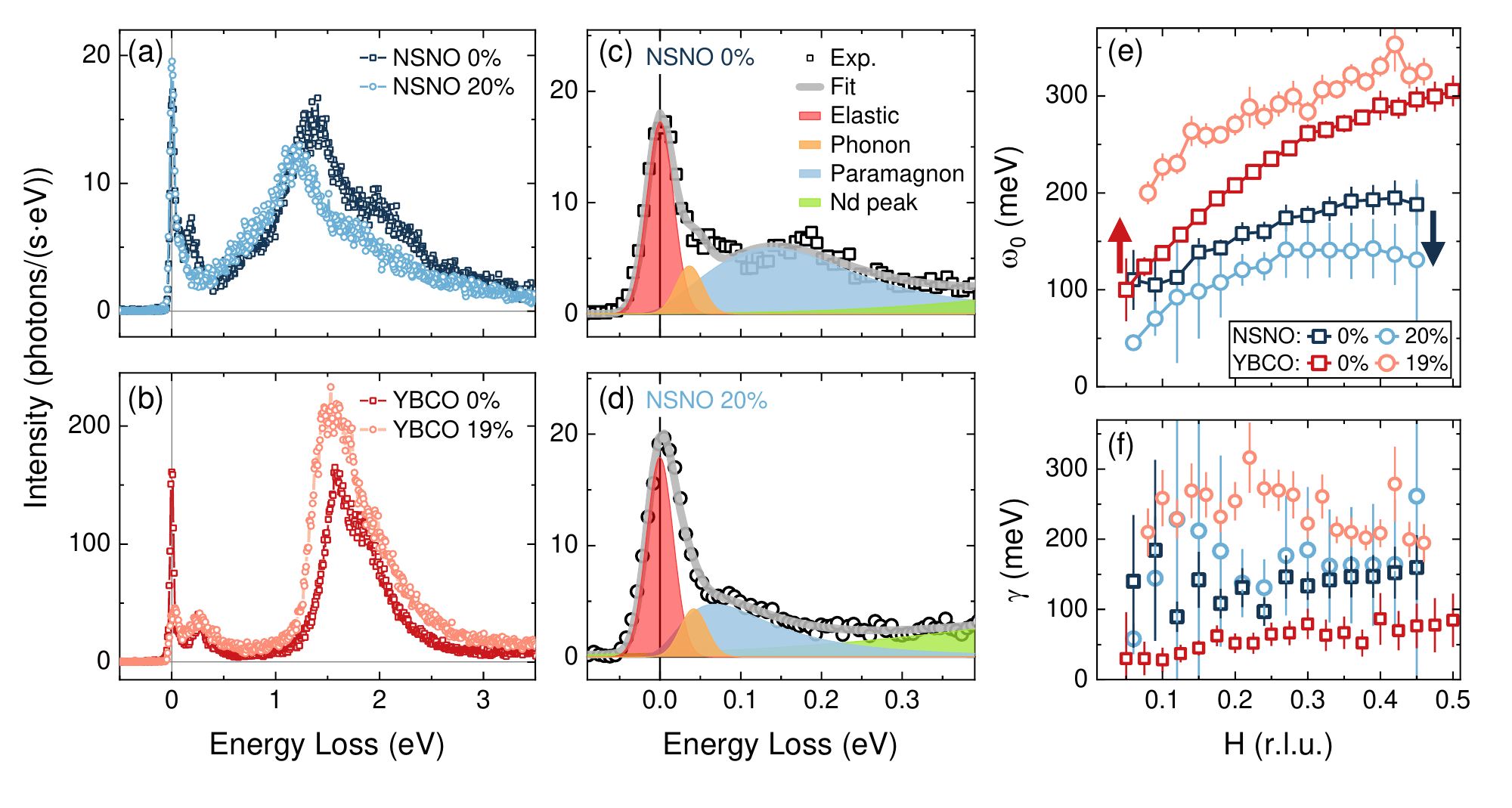}
\caption{\label{fitting} 
Selected RIXS spectra of NSNO at $\mathbf{Q} = (0.36,0)$ (a) and YBCO 0\% at $\mathbf{Q} = (0.35,0)$ and 19\% at $\mathbf{Q} = (0.36,0)$ respectively. (b). Fitting of the low-energy region, using the damped harmonic oscillator model for the paramagnon for NSNO 0\% (c) and 20\% (d). The tail of the Nd hybridization peak (green curve) also includes charge continuum and dd contributes. The momentum dependence of the undamped frequency $\omega_0$ and damping coefficient $\gamma$ are presented in panels (e) and (f) respectively.} 
\end{figure*}

\emph{Experimental details} -- RIXS spectra were measured at the beamline ID32 of the  European Synchrotron (ESRF) in Grenoble, France \cite{Brookes}. The combined resolution at the Ni L$_3$ edge (854.3\,eV) was 39\,meV, while at Cu L$_3$ (931\,eV) it was 30 and 40\,meV for spectra without and with polarization analysis respectively. The scattering angle $2\theta$ was fixed at 149.5$^\circ$ and the incident radiation was linearly polarized in the horizontal scattering plane ($\pi$ polarization), so to enhance the spin-flip excitations for positive values of the momentum transfer, i.e., for grazing emission geometry. Momentum-resolved measurements were carried out on both samples along the (H,0) direction of the Brillouin zone. The spin-flip nature of the various features was assessed by measuring the two linear polarization components $\pi'$ and $\sigma'$ of the scattered radiation thanks to the polarimeter installed on the ERIXS spectrometer \cite{Brookes, braicovich2014simultaneous}. The temperature was set to 20\,K in order to minimize anti-Stokes signal and radiation damage of the samples.
We performed measurements on 9\,nm-thick films of Nd$_{1-x}$Sr$_{x}$NiO$_2$ (hereafter NSNO), with doping levels of $x$ = 0 (undoped) and 0.2 (optimally doped). Both samples were obtained from pristine Nd$_{1-x}$Sr$_{x}$NiO$_3$ perovskite films grown by Pulsed Laser Deposition (PLD) on SrTiO$_3$(001) (STO) single crystals substrate, and capped with three unit cells of STO. The perovskite was then reduced to the IL structure via a soft-chemistry process as described in ref.\,\onlinecite{Krieger2023growth}. The sample preparation was carried out at IPCMS, Strasbourg, France. As cuprate reference, we used 
YBa$_2$Cu$_3$O$_{7-\delta}$: 
one 150\,nm-thick film was grown on a (110) (LaAlO$_3$)$_{0.3}$(Sr$_2$TaAlO$_6$)$_{0.7}$ (LSAT) substrate; this sample was almost perfectly undoped with $\delta \simeq 1$ ($p \simeq 0.03$, referred to hereafter as YBCO 0\%). 
Another 100 nm-thick film, grown on (001) STO,  was fully oxygenated, with $T_c \simeq 88$\,K and $\delta = 0$ ($p = 0.19$, YBCO 19\%). Polarimetric measurements were performed on the same YBCO 19\% at $T$ = 80 K and on another 100\,nm-thick, (001) STO-grown underdoped YBCO film, with $T_c \simeq 63$\,K and $\delta \simeq 0.3$ ($p \simeq 0.13$, YBCO 13\%). The latter was measured at \textit{T} = 60 K. 
YBCO samples were grown by PLD at Chalmers University, Sweden \cite{arpaia2018probing}.


\emph{Results and discussion} -- Fig.\,\ref{NSNO_maps} reports the energy/momentum RIXS intensity maps of NSNO 0\% and NSNO 20\% along the (H,0) in-plane direction of the reciprocal space. 
The undoped sample shows a peak dispersing up to $\sim$200\,meV, which is assigned to a spin wave excitation (magnon) \cite{lu2021magnetic,krieger2022charge}, marked by the light-blue line as a  guide to the eye. The feature is much broader for the NSNO 20\% so that the dispersion is harder to visualize. This effect can be attributed, as in cuprates \cite{PengPRB}, to magnetic disorder in the doped samples \cite{lu2021magnetic,krieger2022charge}. 
In Figures\,\ref{fitting}(a)-(b) we compare RIXS spectra of NSNO and YBCO at the same doping level. We can notice the broadening and overall softening of the orbital excitations (between 0.5\,eV and 3\,eV) upon doping, consistently with previous observations \cite{rossi2021orbital,fumagalli2019polarization}. In YBCO the shift of the $dd$ peaks looks smaller  than in the IL nickelate, but with a larger broadening. We can explain this by the different nature of the doping holes: in cuprates, the relatively small value of the charge-transfer $\Delta$ drives them into the oxygen band, without affecting significantly the energy of copper orbitals. Conversely, in NSNO the positive carriers are added to the Ni $3d$ states, thus contributing to the downward shift of the Fermi level on Ni bands \cite{rossi2021orbital,hepting2020electronic}. 

In order to compare quantitatively the spin excitations in the four samples, we fitted the spectra with the damped harmonic oscillator (DHO) model previously used for cuprates \cite{PengPRB} and IL nickelates \cite{lu2021magnetic}. As shown in Fig.\,\ref{fitting}(c-d), we fitted also the elastic peak and one phonon with resolution-wide Gaussian function, and a very broad Gaussian for the tail of the high-energy electronic excitations. The momentum dependence of the undamped energy $\omega_0$ and damping coefficient $\gamma$ of the magnon/paramagnon are shown in panels (e-f). 

The dispersion curves of magnetic excitations exhibit a similar shape across all samples. However, in the infinite-layer (IL) nickelates, the curve is located at approximately half the energy of the cuprates, indicating a smaller nearest-neighbor superexchange interaction \cite{lu2021magnetic}. Additionally, the magnetic peak evolves differently with doping in these materials: in IL nickelates, the energy decreases with doping, whereas in cuprates it increases, consistently with previous findings \cite{PengPRB,lu2021magnetic}. Furthermore, the damping grows with doping in YBCO, while it remains unchanged within the statistical confidence range in NSNO. 
In cuprates the increase of the spin excitation energy upon hole doping is attributed to the three-site term enabled by neighboring doped sites, which overcompensates the reduction of the average spin moment \cite{jia2014Persistenta}, and the increase in the damping is justified in terms of shortening of the spin-spin correlation function upon doping (increased spin disorder). How can we explain the different behavior of IL-nickelates? 
With the aim of solving these shortcomings, we decided to go beyond the DHO by measuring more accurately the spectral shape of the spin excitations and by analysing it with a more advanced model.



\begin{figure*}
\centering
\includegraphics[width=\textwidth]{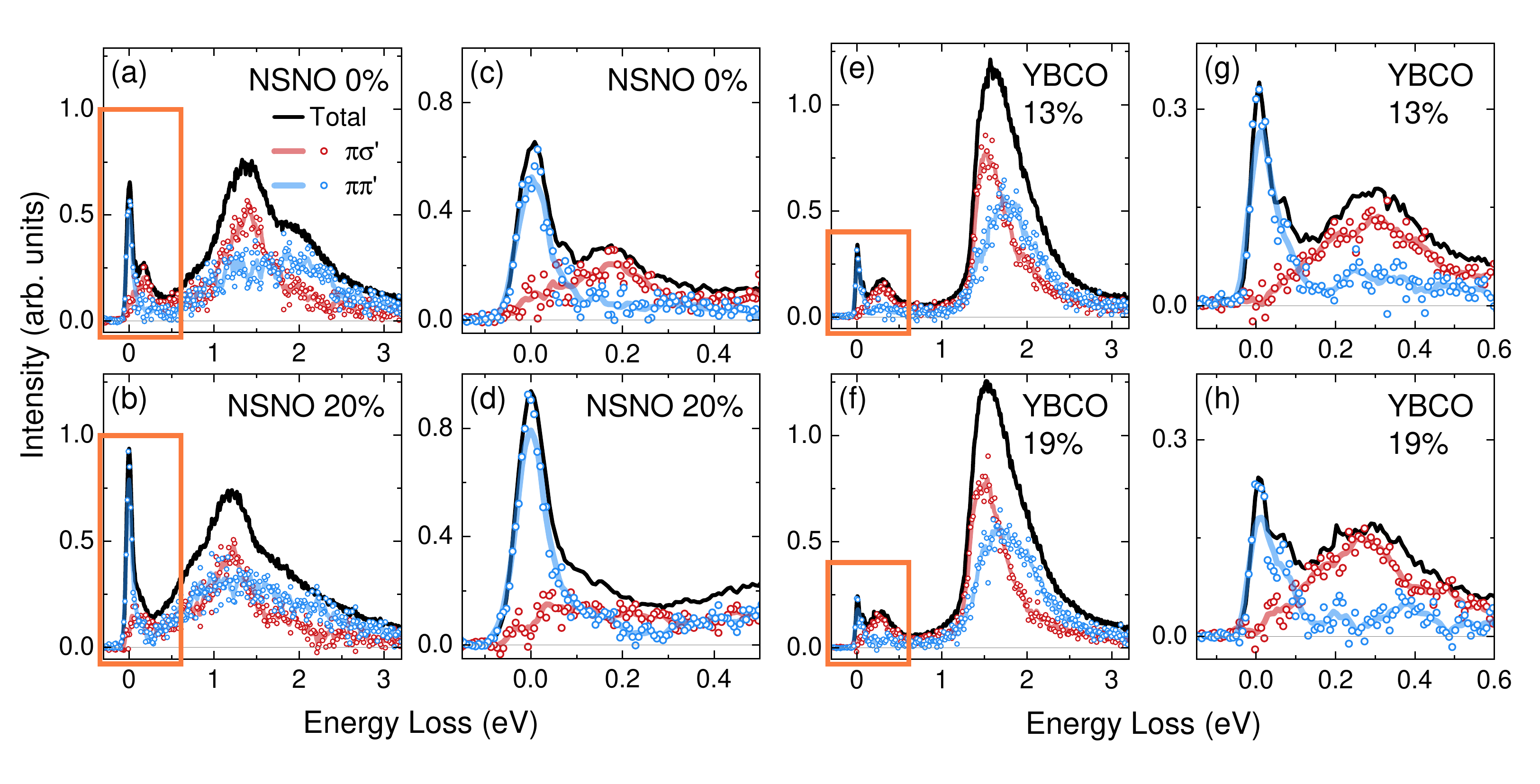}
\caption{\label{polarimeter_spectra} Polarization-resolved spectra of NSNO (a)-(d) at $\mathbf{Q}=(0.36,0)$, and YBCO (e)-(h) at $\mathbf{Q} = (0.44,0)$. The solid black line is the average of the crossed polarization ($\pi\sigma'$, red dots) and parallel polarization spectra ($\pi\pi'$, blue dots). The solid blue and red lines show spectra smoothings over 7 points.}
\end{figure*}

\emph{Polarimeter analysis} -- The broadness of the spin excitations in the doped sample makes it hard to single-out pure spin-flip excitations from the spin-conserving ones (e.g. bimagnons).  Single-magnon excitations have pure crossed polarization character, because they imply the transfer of one unit of angular momentum from the scattering photon to the sample, i.e., a 90° rotation of the linear polarization of the x-ray photon \cite{braicovich2010momentum,jia2014Persistenta,ament2009theoretical,haverkort2010theory}. Therefore, we performed a polarization analysis of the scattered x-rays \cite{fumagalli2019polarization, Moretti}, which is shown in Fig.\,\ref{polarimeter_spectra}.
For NSNO 0\% and YBCO 13\% and 19\%, the 200-300 meV peaks linked to magnetic excitations have almost pure crossed character ($\pi,\sigma'$), while the intensity of the parallel polarization spectrum ($\pi,\pi'$), mostly due to spin-conserving charge excitations (electron-hole pairs) and to bimagnons, is much weaker and featureless.
The $dd$ excitations exhibit different trends well described by single-ion model calculations \cite{fumagalli2019polarization, Moretti}. The polarimetric analysis allows us to extract the pure spectral shape of the spin excitations, otherwise strongly mixed with the tail of the elastic and phonon peak in the unpolarized spectrum, thus enabling a more accurate analysis with theoretical models. 

\emph{Microscopic model of magnetic excitations} -- Having obtained the pure spin part of the mid-infrared range spectrum, we carry out a microscopic analysis by employing a layered square-lattice Hamiltonian $\hat{\mathcal{H}} \equiv \sum_{i \neq j} t_{ij} \hat{a}_{i\sigma}^\dagger \hat{a}_{j\sigma} + U \sum_i \hat{n}_{i\uparrow} \hat{n}_{i\downarrow} + \frac{1}{2}\sum_{i \neq j} V_{ij} \hat{n}_i \hat{n}_j$, extending two-dimensional Hubbard model recently proposed as a unifying framework describing low-energy electronic states of both Cu and Ni based superconductors \cite{kitatani2020nickelate}. To address multi- and infinite-layer systems, we select equal in-plane and out-of-plane lattice constants. The model is analyzed using the variational wave function approach combined with $1/\mathcal{N}_f$ expansion (VWF+$1/\mathcal{N}_f$) \cite{Fidrysiak2021Method,fidrysiak2021unified}, see Supplemental Material (SM) \cite{SuppMat}. The free microscopic parameters are the on-site Coulomb repulsion $U$, the in-plane nearest- and next-nearest-neighbor hopping integrals ($t$ and $t^\prime$), the screened intersite Coulomb interactions $V_{ij}$, and the hole-doping $\delta$. From DFT + DMFT (Dynamical Mean Field Theory) calculations the estimated values of $t$ and $t^\prime$ are comparable in cuprates and IL-nickelates \cite{kitatani2020nickelate}, whereas the effective value of $U$ is expected to differ by a factor $\sim 2$ as it is constrained by the experimental paramagnon bandwidth $\propto \frac{4 t^2}{U}$. Microscopically, such variation of $U$ has been interpreted in terms of distinct one-band model mapping for charge-transfer and Mott-Hubbard insulators \cite{been2021electronic,Feiner1996CTI_vs_MottHubbard}. Hereafter we refer to $U/|t| \sim 6$-$8$ and $U/|t| \gtrsim 9$ as cuprate- and nickelate regimes, respectively. The role of $V_{ij}$ terms is to suppress phase-separation, and they are determined as a solution of generalized lattice Poisson equation \cite{SuppMat}.

\begin{figure}
\centering
\includegraphics[width=1.03\columnwidth]{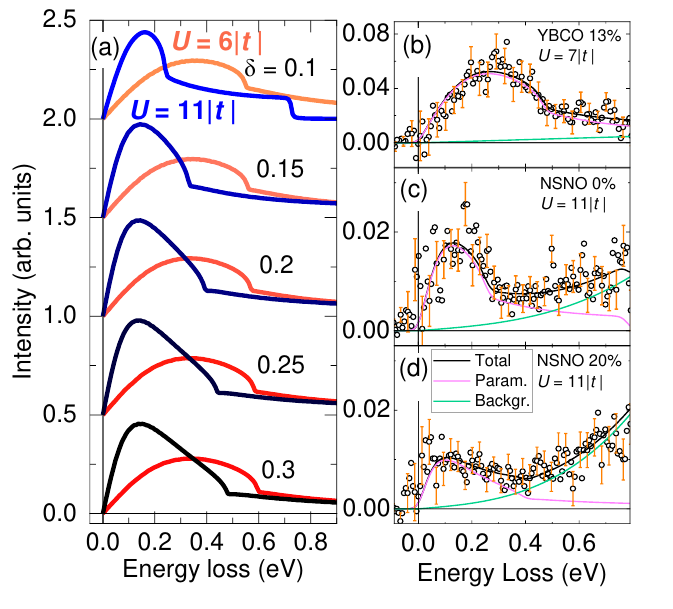}
\caption{\label{new_model_fit} (a): Theoretical dissipative parts of Hubbard-model susceptibilities $\chi^{\prime\prime}(\mathbf{Q}, \omega)$ at $\mathbf{Q} = (0.36, 0)$ for various doping levels $\delta$; pink-red curves refer to the cuprate regime ($U = 6 |t|$), blue-black curves to nickelate regime ($U = 11 |t|$). The remaining parameters are: $t = -0.4\,\mathrm{eV}$, $t^\prime = 0.25|t|$, $k_B T = 0.35 |t|$. Theoretical susceptibilities were convolved with a 39\,meV FWHM Gaussian to account for experimental resolution. (b): Low-energy fit of YBCO 13\% cross-polarized spectrum (empty dots) at $\mathbf{Q} = (0.44,0)$ using our magnetic susceptibility (pink curves) plus a cubic background (aquamarine). (c)-(d): Same for NSNO 0\% and 20\% respectively, at $\mathbf{Q} = (0.36,0)$. Microscopic parameters are reported in Table~\ref{tab:fitparam} for the two regimes.}
\end{figure}

In Fig.\,\ref{new_model_fit}(a) we compare the hole-doping evolution of the calculated dissipative part of dynamical spin susceptibility $\chi_s^{\prime\prime}(\omega, \mathbf{Q})$ at $\mathbf{Q} = (H,K) = (0.36, 0)$ r.l.u. for $U = 6 |t|$ (red) and $U = 11 |t|$ (blue),  both with the same hopping integrals ($t = -0.4\,\mathrm{eV}$, $t^\prime = 0.25 |t|$) and non-local Coulomb terms. The temperature is set to $k_B T = 0.35 |t| \sim 0.12\,\,\mathrm{eV}$ so as to stabilize the paramagnetic state and preserve the physical hierarchy $k_B T \lesssim \omega_0, \gamma$. Note that the simulated spectra separate into a resonant low-energy paramagnon peak and a featureless particle-hole continuum, extending to larger energies, which goes beyond the DHO-model phenomenology. The $U = 6 |t|$ solution contains a robust paramagnon centered at $\approx 0.3 \,\,\mathrm{eV}$, in agreement with previous theoretical studies of magnetic excitations in doped cuprates \cite{fidrysiak2021unified,nilsson2019dynamically,fidrysiak2020robust}. On the other hand, in the nickelate regime (blue curves) the paramagnons \emph{are not} robust and we observe a crossover from a well-defined peak at $\delta = 10\%$ to a broader asymmetric lineshape at $\delta \gtrsim 20\%$. \textit{This suggests that the magnitude of the effective $U$ is the principal microscopic factor differentiating the observed doping evolutions of the magnetic excitations.}

\begin{table}
     \begin{tabular}{m{2.4cm} m{2.4cm} m{2.4cm}}
     \hline
     \hline
      Parameter &   Cuprate    & Nickelate \\
      \hline
       $t$      &   -0.35 eV   & -0.4 eV  \\
       $U$      &    7$|t|$ = 2.45 eV    & 11$|t|$ = 4.4 eV \\
       $t'$   &    0.25$|t|$      & 0.25$|t|$ \\
       $k_BT$   &    0.35$|t|$ & 0.3$|t|$  \\
       $\delta$ &    0.13      & 0.1, 0.2 \\
       \hline
       \hline
    \end{tabular}
    \caption{Microscopic Hubbard parameters obtained by fitting the polarimetric experimental data of YBCO 13\%, NSNO 0\% and NSNO 20\% with the theoretical dynamical spin susceptibility.}
    \label{tab:fitparam}
\end{table}

We fitted the polarimetric cross-polarized spectra of YBCO 13\%, NSNO 0\% and NSNO 20\% with the theoretical dynamical spin susceptibilities (Fig.\,\ref{new_model_fit}(b)-(d)), arriving at the parameters listed in Table\,\ref{tab:fitparam}. Details of the microscopic-model analysis are presented in the SM \cite{SuppMat}. The Mott self-energy for nickelates $U \sim 4.4 \, \mathrm{eV}$ falls in between metallic nickel and NiO \cite{been2021electronic} and is in agreement with previous works \cite{worm2022correlations, kitatani2020nickelate, been2021electronic, pavarini2001band}. To account for self-doping, we assume a nonlinear dependence of $\delta$ on the Sr concentration $x$ in NSNO, as suggested by recent DFT+DMFT calculations \cite{kitatani2020nickelate}, namely $\delta = 0.1$ for NSNO 0\%,  and $\delta = 0.2$ for NSNO $20\%$.  
We emphasize that for the two nickelate dopings all parameters are identical except $\delta$ and that  the same value of $t^\prime/t$ can be used for YBCO and NSNO data. The latter finding is particularly relevant because $t^\prime/t$ is known to heavily affect the shape of the susceptibility and because the literature is not univocal on this point, with some authors proposing higher values of $t^\prime/t$ $\sim$ 0.3 - 0.4 \cite{pavarini2001band, botana2020similarities} and others using values comparable to ours \cite{nomura2019formation,been2021electronic,kitatani2020nickelate}. We notice that fitting the spectra without polarization analysis would lead to smaller $t^\prime/t$ $\sim$ 0.15, see Fig.\,2 of SM \cite{SuppMat}. The crucial role played by polarization-resolved analysis in this discernment is apparent. It is proved that the magnon softening in nickelates comes as a consequence of the smaller $t/U$ ratio (see Fig.\,\ref{new_model_fit} (a)), giving further support to our model with no need of involving spin dilution effects, as previously proposed \cite{lu2021magnetic}.

\emph{Conclusions and perspectives} -- By comparing the RIXS spectra of IL nickelates (NSNO) and cuprates (YBCO), both in undoped and doped states, we have highlighted the distinct doping dependence of magnetic excitations in these two families of quantum materials, in terms of both energy and damping. To further investigate these differences, the polarization-resolved RIXS spectra of NSNO revealed two key findings:  i) the presence of a significant  non-crossed component, detectable even in the nominally undoped sample, likely due to self-doping; and ii) the accurate shape of the pure magnetic response, which can be properly fitted with theoretical models going beyond the commonly used DHO model. Using our Hubbard-based approach, we successfully reproduced the different evolution of the dynamic spin response with doping in cuprates and nickelates, yielding reliable estimates of crucial parameters, such as the Mott-Hubbard $U$ ($\sim 2.45$\,eV in YBCO and $\sim 4.4$\,eV in NSNO). 

In this context, the role of the charge transfer $\Delta$ remains to be clarified. 
As cuprates and IL nickelates are  charge-transfer ($\Delta < U$) and Mott-Hubbard ($\Delta > U$) insulators, respectively, $\Delta$ is expected to have significantly different values between the two materials. 
Such an hypothesis calls for further investigation, starting from the expected decrease of the exchange coupling $J$ when $\Delta$ is increased \cite{wang2018magnon}. 

\begin{acknowledgments}
F.R., L.M., M.M.S., M.S. and G.G. acknowledge support by the projects PRIN2017 ``Quantum-2D” ID 2017Z8TS5B and PRIN 2020 "QT-FLUO" ID 20207ZXT4Z of the Ministry for University and Research (MIUR) of Italy. R.A. acknowledges support by the Swedish Research Council (VR) under the Project 2020-04945. M.F. acknowledges support by grants Miniatura No. DEC-2021/05/X/ST3/00666 and Opus No. UMO-2021/41/B/ST3/04070 from Narodowe Centrum Nauki (NCN). D.P. acknowledges support from the project ANR-JCJC FOXIES (ANR-21-CE08-0021). For the purpose of Open Access, the author has applied a CC-BY public copyright licence to any Author Accepted Manuscript (AAM) version arising from this submission.
This research used ESRF beam line ID32 under the proposals HC4831, HC5221 and HC5222. This work was performed in part at Myfab Chalmers.
\end{acknowledgments}

\bibliography{Biblio}
\end{document}